\documentclass[final,3p]{elsarticle}
\biboptions{numbers,sort&compress}

\usepackage{amssymb}
\usepackage{lineno}
\usepackage[]{siunitx}
\usepackage{multirow}
\usepackage{wasysym}
\usepackage{xcolor}
\usepackage{subfigure}
\usepackage[utf8]{inputenc}
\usepackage{makecell}
\usepackage[page,toc,titletoc,title]{appendix}
\usepackage[subfigure]{tocloft}
\usepackage[section]{placeins} 

\usepackage{siunitx} 
\DeclareSIUnit\sq{\ensuremath{\Box}}                           

\usepackage[rawfloats=true]{floatrow}

\usepackage{xcolor}

\usepackage{hyperref}

\journal{Nucl. Instrum. Meth. A}

\begin{document}

\begin{frontmatter}



\title{Design and performance of the Fermilab Constant Fraction Discriminator ASIC}


\author[a,b]{Si Xie}\ead{sxie@fnal.gov}
\author[a]{Artur Apresyan}
\author[a,c]{Ryan Heller}
\author[a]{Christopher Madrid}
\author[a]{Irene Dutta}
\author[d]{Aram Hayrapetyan}
\author[a]{Sergey Los}
\author[a]{Cristi\'an Pe\~na}
\author[a]{Tom Zimmerman}

\cortext[1]{Corresponding author}

\address[a]{Fermi National Accelerator Laboratory, PO Box 500, Batavia IL 60510-5011, USA}
\address[b]{California Institute of Technology, Pasadena, CA, USA}
\address[c]{Lawrence Berkeley National Laboratory, Berkeley, CA, USA}
\address[d]{A.~Alikhanyan National Science Laboratory, Yerevan, Armenia}

\begin{abstract}
We present the design and performance characterization results of the novel Fermilab Constant Fraction Discriminator ASIC (FCFD) developed to readout low gain avalanche detector (LGAD) signals by directly using a constant fraction discriminator (CFD) to measure signal arrival time. 
Silicon detectors with time resolutions less than \SI{30}{\ps} will play a critical role in future collider experiments, and LGADs have been demonstrated to provide the required time resolution and radiation tolerance for many such applications.
The FCFD has a specially designed discriminator that is robust against amplitude variations of the signal from the LGAD that normally requires an additional correction step when using a traditional leading edge discriminator based measurement.
The application of the CFD directly in the ASIC promises to be more reliable and reduces the complication of timing detectors during their operation. 
We will present a summary of the measured performance of the FCFD for input signals generated by internal charge injection, LGAD signals from an infrared laser, and LGAD signals from minimum-ionizing particles.
The mean time response for a wide range of LGAD signal amplitudes has been measured to vary no more than \SI{15}{\ps}, orders of magnitude more stable than an uncorrected leading edge discriminator based measurement, and effectively removes the need for any additional time-walk correction.
The measured contribution to the time resolution from the FCFD ASIC is also found to be \SI{10}{\ps} for signals with charge above \SI{20}{\femto\coulomb}.
\end{abstract}



\end{frontmatter}

\clearpage
\tableofcontents


\section{Introduction}

Precise timing information will play a critical role in the performance of future tracking detectors and currently poses a profound challenge to their development~\cite{4DTracking_WhitePaper}. 
Tracking detectors capable of achieving 5--25~\si{\ps} timing resolution and 5--30~\si{\micro\m} position resolution are needed for many proposed future colliders including the Muon Collider~\cite{accettura2023muon}, the FCC-hh~\cite{Sickling, Wulz277931111}, and the Electron–Ion Collider (EIC)~\cite{AbdulKhalek:2021gbh}.
Low gain avalanche detectors (LGAD) are a leading candidate for such applications~\cite{ACLGADprocess, 8846722, RSD_NIM, firstAC}, and have been demonstrated to achieve performance of 20--30~\si{\ps}~\cite{Apresyan:2020ipp,Heller_2022,Madrid:2022rqw,OTT2023167541,TORNAGO2021165319}.
Design of the front-end electronics capable of extracting precision timing information from LGAD sensors presents many challenges, and plays a key role in the applications of the LGAD technology.  
We will present the design and characterization of a new application--specific integrated circuit (ASIC) called the Fermilab constant fraction discriminator (FCFD) designed for time-stamping LGAD signals.
The initial version of the FCFD, referred throughout the paper as FCFDv0, provides an analog discriminator signal for a single channel.

Efficient hit detection and excellent timing is the main task of the front-end electronics. 
In order to achieve the desired timing resolution, various approaches have been taken. 
Traditionally, time-of-flight (TOF) detectors use a discriminator in combination with time-to-digital converters (TDCs) to measure the Time-of-Arrival (ToA) and Time-over-Threshold (ToT), such as implemented in the readout ASICs for the CMS and ATLAS timing detectors~\cite{ETROC1, ALTIROC2}. 
The ToT or signal amplitude is critically needed in order to correct for the time-walk effect, in which the ToA measurement changes sharply or ``walks" depending on the amplitude of a signal pulse. 
The time-walk dependence can change with sensor bias voltage, temperature, or total irradiation dose due to their impact on the signal amplitude. 
Therefore, a time-walk correction needs to be applied to each channel separately in order to achieve the optimal timing resolution and continuously be updated throughout the lifetime of the detector. 

The solution developed in this paper utilizes an approach that does not require offline corrections or calibrations and is much simpler in operation.  
We have previously performed in-depth simulation studies comparing the performance of various algorithms to time-stamp signals from LGAD sensors, where we demonstrated that the CFD approach achieves better performance, especially for low signal-to-noise (S/N) systems~\cite{CFDSim}. 
This enables simple and robust timing measurements of LGAD signals that vary in amplitude by at least a factor of 10, with no critical threshold setting or corrections required. 
The FCFDv0 uses several new techniques to achieve low power, area, jitter, time-walk, and drift. 
In this paper, we present the analog front-end of an ASIC implemented in TSMC 65~nm CMOS technology node. 
This chip is the first prototype to study and optimize the performance of CFD-approach to LGAD time-stamping. 

\section{LGAD sensors general signal characteristics}\label{sec:sensor}

LGAD sensors are produced by introducing an additional layer of p$^+$ material, such as Boron, close to the n-p junction of traditional silicon sensors. 
This results in a very high electric field in the region within a depth of a few micrometers of the junction. 
This region is referred to as the ``gain'' or ``multiplication'' layer. 
The initial signal generated by the ionization of the minimum-ionizing particle as it traverses the sensor material is amplified through an avalanche process initiated by electrons passing through the gain region. 
The amplified signals maintain fast slew rate, resulting in excellent timing characteristics~\cite{Apresyan:2020ipp,Heller_2022,Madrid:2022rqw,TORNAGO2021165319,OTT2023167541}. 
Signals are read out from the n$^+$ cathode, and since the bulk material is a high resistivity p-type silicon, a shallow uniform p-spray doping is usually implemented to isolate the cathodes. 
To reduce the magnitude of the electric field at the perimeter of each signal pad, an additional deep n+ doping region called the Junction-Terminating Extension (JTE) is implemented. 
JTEs are s characteristic feature of LGAD sensors, and results in regions of no gain between pads referred to as the ``inter-pad gap''. 

The FCFDv0 chip design specifications were optimized targeting 50~$\mu$m thick LGAD sensors with 1.3$\times$1.3 mm$^2$ surface area. 
The capacitance and the rise time of the signal input to the front-end electronics are 3.4~pF and 500~ps, respectively, which are key parameters used to optimize the front-end amplifier. 
The most probable signal charges from LGADs ranges between 5 and 35~fC, depending on the bias voltage applied to the sensors. 
The LGAD sensors can attain gains between 10 and 20 at relatively low bias voltage.
With increasing radiation damage accumulated through operation in a high particle flux environment, the bias voltage must be increased to maintain high gain~\cite{Mazza_2020}. 
The intrinsic time resolution of the LGAD can be maintained around 30 ps even at neutron fluences up to $1\times$10$^{15}/\mathrm{cm}^{2}$.
The design requirement for the FCFD chip is to remain a subdominant contributor to the overall timing resolution.

\begin{figure}[htp]
\centering
\includegraphics[width=0.9\textwidth]{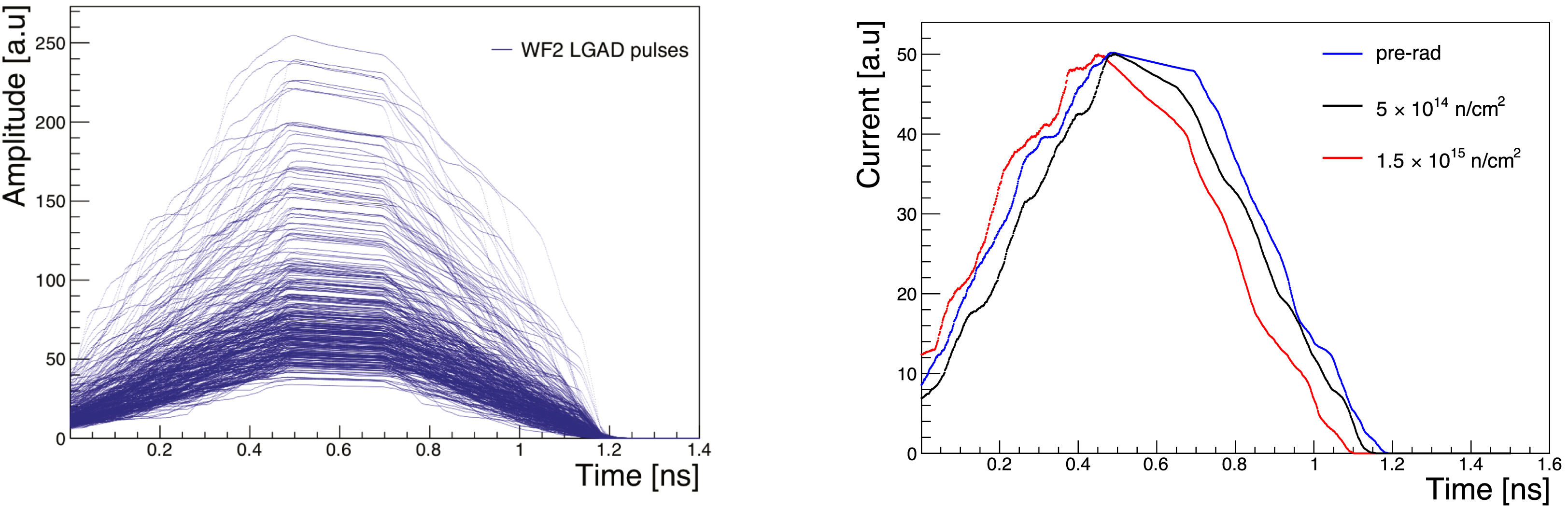}
\caption{One thousand simulated LGAD signal waveforms from WF2 for a pre-radiated sensor (left). Characteristic example LGAD signal waveforms for different irradiation levels (right).
\label{fig:LGADPulses}}
\end{figure}

Systematic studies of the timing performance of LGAD sensors~\cite{CFDSim}, simulated by the 2-dimensional silicon simulator Weightfield2 (WF2)~\cite{CENNA2015149}, and parametric simulation of amplifiers and time-to-digital converters strongly indicated that the performance of a CFD outperforms the leading edge discriminator even with the most ideal of time-walk correction scenarios implemented.
Figure~\ref{fig:LGADPulses} shows the unprocessed LGAD output current from WF2 for 1,000 signal pulses (left) and characteristic example pulses for all the irradiation levels studied (right). 
The performance advantages of the CFD is particularly strong for LGAD sensors after irradiation exceeding $5\times10^{14}$ in neutron fluence. 
Realistic granularity of the time and amplitude bins necessary for the time-walk correction present major technical and cost demands on maintaining accuracy of the time-walk correction.
Implementation of complicated calibration and post-processing schemes necessary for maintaining time-walk correction accuracy also presents major operational challenges.

The same simulated waveforms were used to optimize the design of the FCDFv0 ASIC.
Moreover, the charge injection circuit of the FCDFv0 ASIC was designed to closely mimic the simulated LGAD waveforms, so that benchtop tests of the ASIC performance using the injected signal pulses can be as realistic as possible. 

\section{Design of the FCFD}\label{sec:design}

\begin{figure}[htp]
\centering
\includegraphics[width=0.8\textwidth]{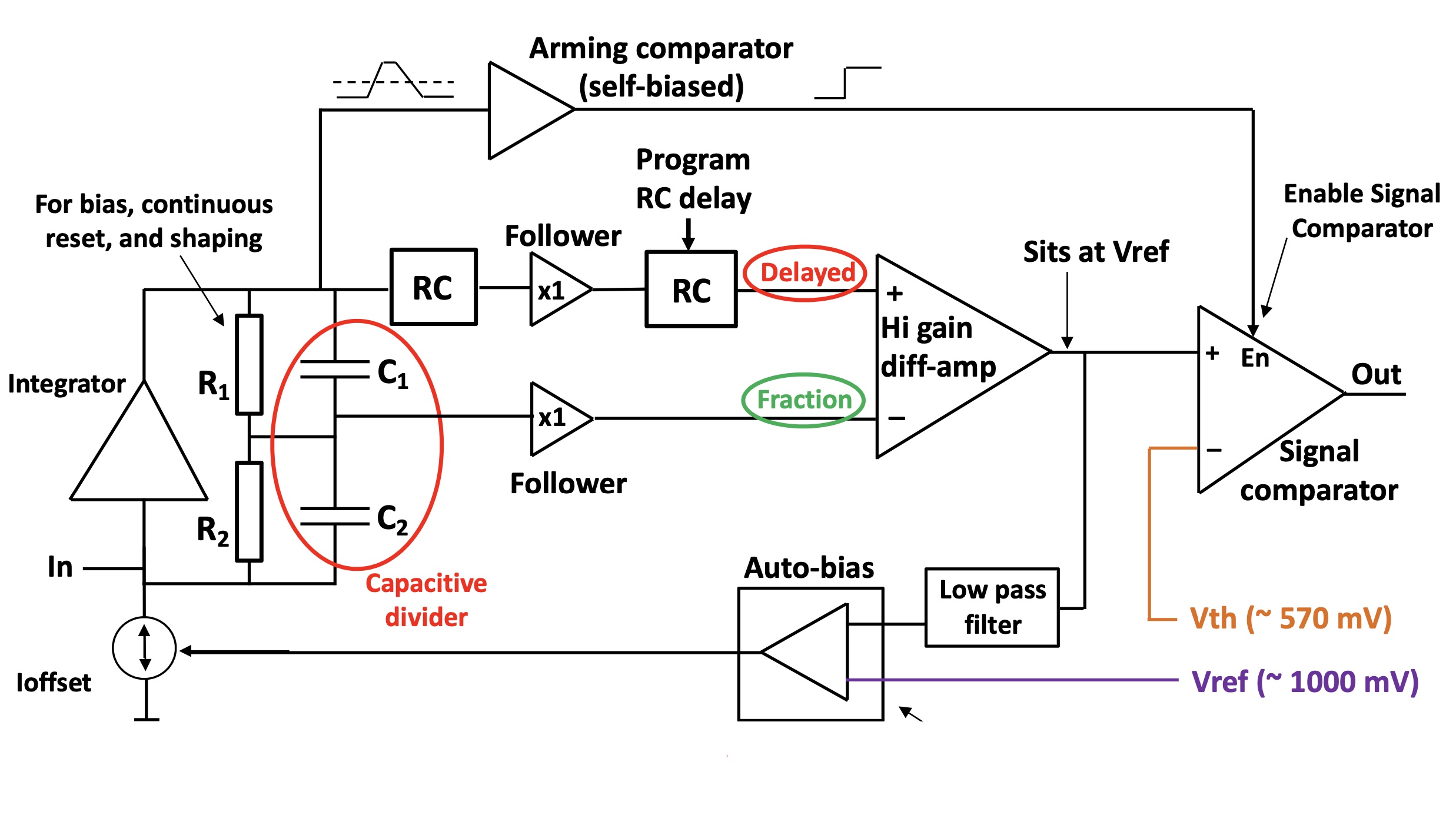}
\caption{A simplified diagram of the FCFDv0 ASIC, and its main blocks.
\label{fig:FCFDv0}}
\end{figure}

Following the guidance from the simulation studies, we developed the first version, FCFDv0, of the CFD-based readout chip optimized for LGAD sensors. 
We present a detailed description of the main components of the FCFDv0 ASIC : the Integrator, the Follower, the Arming Comparator and the Auto-bias. 
Figure~\ref{fig:FCFDv0} shows a simplified diagram of the ASIC and labels components referenced in the following section.

\textbf{Integrator}: The function of the Integrator is to turn the signal charge pulse into a voltage form in order to perform the constant fraction measurements. 
The integrator is built from a single-stage common source amplifier with regulated cascode on the input transistor, and cascode current source, to achieve high open loop gain. 
Capacitive feedback with two capacitors in series performs signal integration while also generating the fractional signal. 
Each capacitor has a parallel resistance in order to provide some shaping and continuous reset. 
The equivalent capacitance of series of C1 and C2 determines the gain at the integrator output (“charge transfer gain”), and their values are chosen by the expected dynamic range for these sensors (5-35 fC). 
The values of C1 and C2 are approximately equal to each other to achieve approximately $50\%$ signal fraction at the comparator. 
The R1 and R2 resistances provide some shaping and continuous reset, and their values are chosen such that this signal completely resets after \SI{25}{\ns} interval. 
One of the novel features is the introduction of the series feedback capacitors to achieve fractional signals. 
In order for this to work best, one needs to minimize the parasitic capacitances at the junction of series capacitors. 
The layout of the series capacitors was done carefully to avoid parasitics to substrate, and we also used our specially designed follower~\cite{Follower} which has a gain of nearly one and extremely low input capacitance. 
The output of the follower was used to drive the metal plate that sits under the two capacitors in order to bootstrap the junction of the two-capacitor node, effectively removes parasitic capacitance on that node. 

\textbf{Follower}: A simple voltage follower typically has gain of less than one due to two reasons: 1) the bulk is held at constant potential, and 2) the drain is held at constant potential. 
In order to achieve gain of 1, both the bulk and the drain need to move the same amount in voltage as the source does. 
A simple and elegant way to achieve this is to use transistors of different threshold voltages so that the HVT is the follower, and the LVT drives the drain of the follower~\cite{Follower}. 
This ensures that there is big enough voltage bias from the follower source to the drain to operate as a follower, and that the drain of the follower will very closely follow the source. 
In this configuration any voltage change at the follower input will be seen on all terminals of the follower, thereby making its gain very nearly one, and its input capacitance very nearly zero. 

\begin{figure}[htp]
\centering
\includegraphics[width=0.6\textwidth]{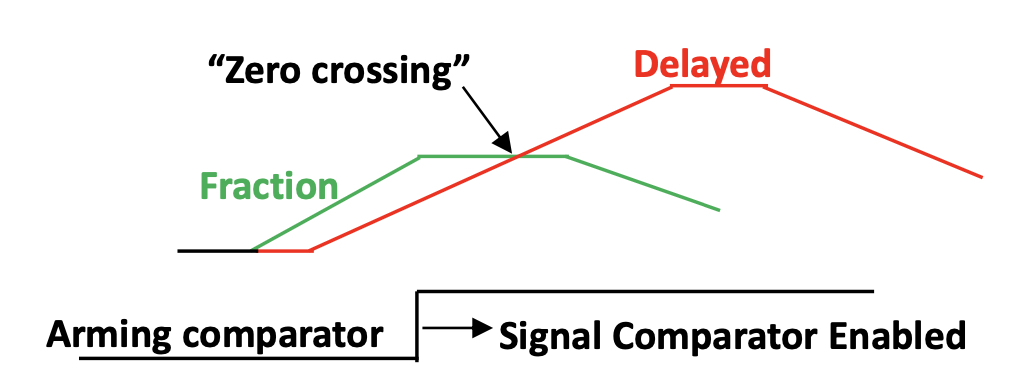}
\caption{Diagram showing the Zero Crossing Comparator
\label{fig:ZCC}}
\end{figure}

\textbf{Arming comparator}: The function of the arming comparator is to inhibit noise triggers at the signal comparator, when there is no signal present. 
The arming comparator detects any signal of appreciable magnitude well above the noise, and enables the signal comparator so that it will fire when the delayed signal crosses the fractional signal. 
It is not critical when and what threshold the arming comparator fires, as long as it fires well above the noise level and below the minimal signal. 
A schematic diagram of the arming comparator is shown in Figure~\ref{fig:ZCC}.

\textbf{Auto-bias}: 
The CFD discriminator is made from a High Gain Diff-amp that feeds a Signal Comparator, as shown in Figure~\ref{fig:FCFDv0}.  
The Fraction and Delayed signals drive the High Gain Diff-amp, which has a single ended output that slews between ground and a logic high (1.2 V).  
This drives the signal comparator, which is an elementary comparator with a diff-amp input and single-ended output.  
This comparator's threshold is set at approximately the midpoint of the logic level swing.  
In the absence of a signal, the Signal Comparator is disabled by the Arming Comparator.  
When a signal arrives, the Fraction signal goes high first, forcing the High Gain Diff-amp output to a logic low.  
Soon after, the Arming Comparator enables the Signal Comparator, so that when the Delayed signal subsequently crosses the Fraction signal, the Diff-amp output will go high and flip the Signal Comparator.  
With ideal building blocks (infinite gain and bandwidth, no offsets), the Diff-amp output DC bias level should rest at exactly the Signal Comparator threshold, so that this comparator will flip at exactly the time when the Delayed signal crosses the Fraction signal, for all signal amplitudes.  
In reality, these building blocks have random offsets and non-zero delays. 

An additional problem is that the Signal Comparator delay varies depending on the signal amplitude ("time-walk"):  small signals have larger delay than large signals.  
In other words, although in theory a CFD has no time-walk, there is still some time delay dependence on signal amplitude.  
In order to compensate for this, the Diff-amp output bias point is set to a level that is not equal to the Signal Comparator threshold.  
As shown in Figure~\ref{fig:ZCC}, this results in the Diff-amp output crossing the Signal Comparator threshold level sooner for smaller signals and later for larger signals.  
This tends to compensate for the amplitude dependent delay of the the Signal Comparator.  
The bias level is set externally in order to allow optimization of this compensation.  
This external level establishes the internal Diff-amp output bias level through the servo action of an Auto-bias circuit that senses the DC level of the Diff-amp output (through a low-pass filter) and applies the appropriate DC input bias current required to correct for all random offsets.  
Therefore with this method, one externally set DC bias voltage sets the proper internal operating point to correct for random offsets and enable optimal comparator time-walk compensation.

In order to monitor the behavior of the CFD as a function of injected pulses, we added an option in the FCFDv0 chip that allows us to ``spy'' on the analog pulses directly by reading out the signal waveforms with an oscilloscope.  
The spy channel is implemented as a simple source follower that can be switched on or off, which is a very useful feature used to calibrate the relationship between the laser signal intensity and injected signal charge, and to monitor the performance of the CFD under different circumstances. 

\section{Experimental setup}\label{sec:setup}

We characterize the performance of the FCDFv0 chip in four complementary ways, probing slightly different characteristics in each case.
The FCFDv0 chip is equipped with an internal charge injection feature, allowing us to probe its response to signal inputs of known amplitude and timing. 
Signals from an LGAD sensor generated by a picosecond fast laser and a beta particle source are used to characterize the chip's response to typical LGAD signals.
Finally, signals produced by particles from a 120~GeV proton beam impinging on the LGAD sensor are used to characterize the chip's response to minimum-ionizing particles. 

\begin{figure}[htp]
  \begin{center}
  \includegraphics[width=0.6\textwidth]{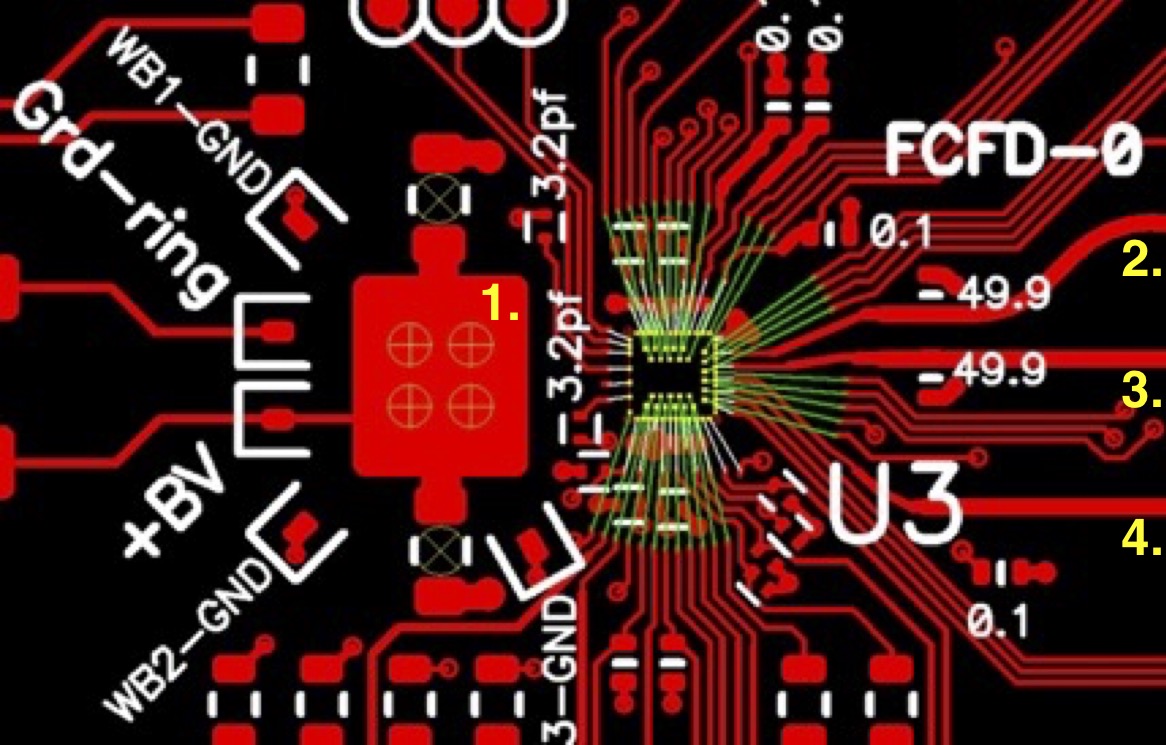}\\
  \caption{Schematic layout of the FCFDv0 readout board, showing the LGAD mounting pad (1) and the impedance-matched microstrip lines (2, 3, and 4). The LGAD mounting pad includes 4 through-holes for tests with beta particles.}
  \label{fig:RBLayout}
  \end{center}
\end{figure}

A specialized readout board, as shown in Figure~\ref{fig:RBLayout}, was designed to evaluate the FCFDv0 performance using an LGAD sensor stimulated by laser, radioactive source, and accelerated particle beam.
The readout board size was chosen to be $128\times101$~mm$^2$ which is a standard size of readout boards routinely used by our group for test beam measurements and allows for the measurements of the FCFDv0 to be performed in a well understood dark box. 
The mounting pad was designed to hold a $2\times2$ pad LGAD over four through-holes aligned with the center of each pad to allow beta rays to pass through unobstructed to a timing reference detector.
The LGAD wire bonding pad structure allows for an equivalent input capacitance connection to the FCFDv0 channels for debugging purposes. In order to minimize wire-bond inductance for the most critical connections, the chip is mounted in a PCB well, flush with the top board surface. Such placement shortens the wire bonds to 0.2-0.25~mm for ground (double wire-bond), 0.4-0.6~mm for fast signal, and 1.3-1.5~mm for power and current bias connections. Microstrip lines are used to rout fast signals, estimated bandwidth for those connections is around 10~GHz. Low voltage power and LGAD bias voltage connections use balun filters to minimize external noise caused by system ground loops. A number of board mounted switches are used to select an amplitude for calibration charge injection, and to configure other FCFDv0 parameters, while an incorporated Pt\_RTD provides monitoring of the board’s temperature.

\begin{figure}[htp]
  \begin{center}
  \includegraphics[width=0.410\textwidth]{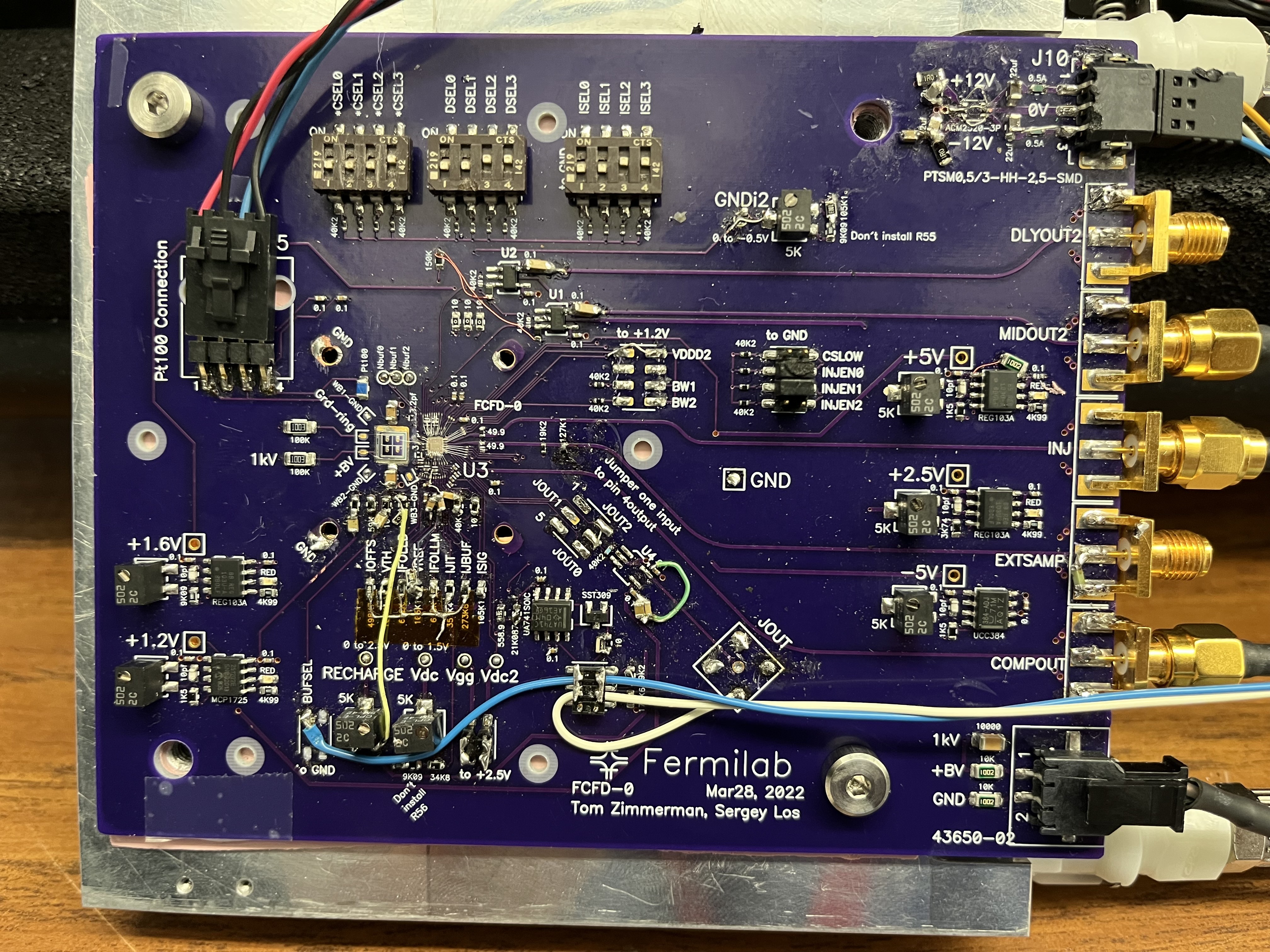}
  \includegraphics[width=0.546\textwidth]{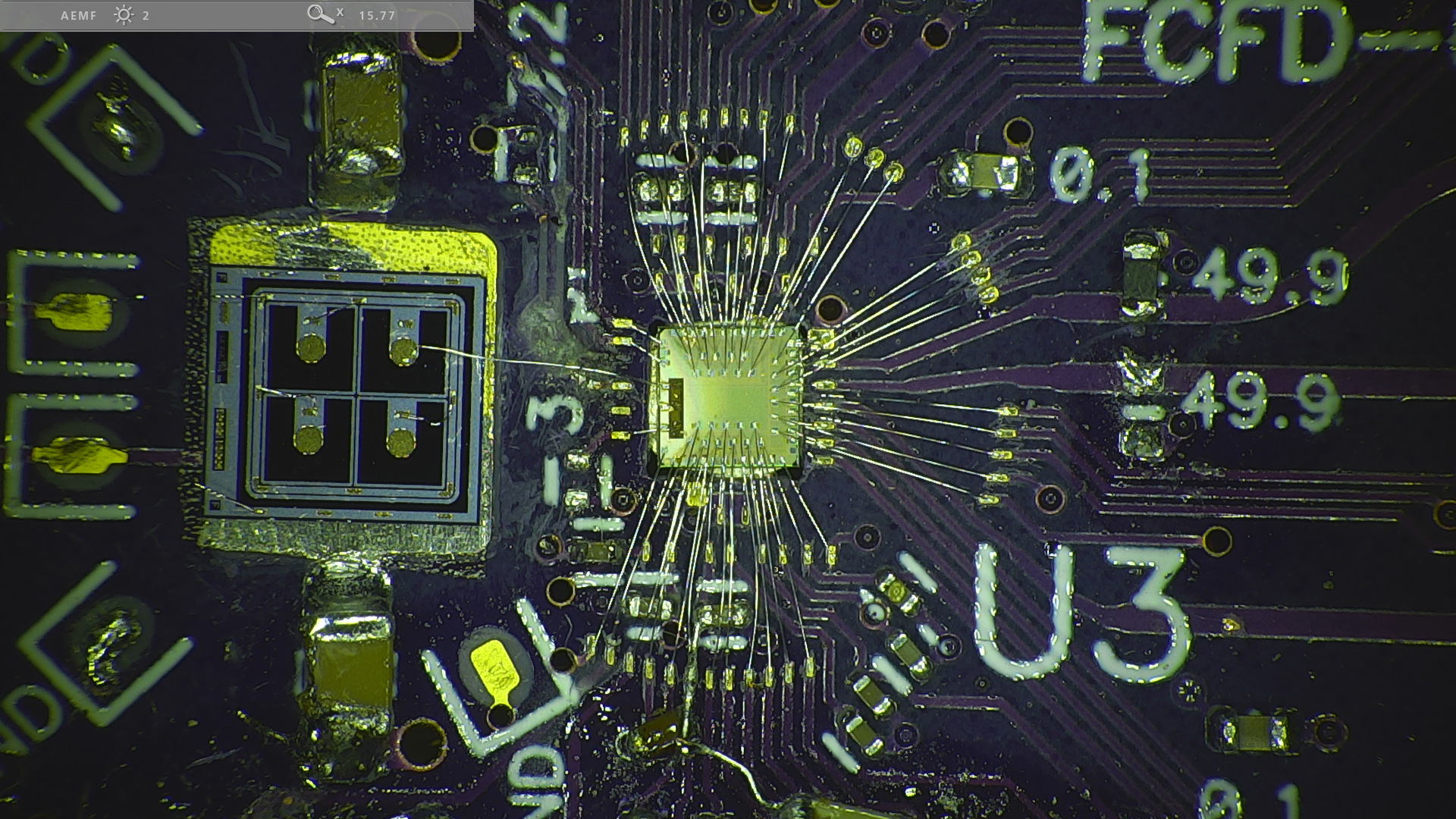}
  \caption{A picture of the FCFDv0 characterization board with an LGAD sensor mounted, showing the switches to select amplitudes for calibration charge injection (left). A closeup picture showing the FCFDv0 wire-bonded to the readout board and to the LGAD sensor (right).}
  \label{fig:RBPic}
  \end{center}
\end{figure}

The FCFDv0 was mounted on the readout board along with the LGAD sensor, as shown in Figure~\ref{fig:RBPic}. The LGAD sensor used in our measurements was tested previously and its characteristics were summarize in Ref.~\cite{HELLER2021165828}. 
The FCFDv0 chip and LGAD mounted readout board were installed in a dark environmental chamber for all measurements. 
This chamber is light-tight and allows for full control of the temperature and humidity to ensure reproducible results. 
Furthermore, the readout board is mounted onto an actively cooled aluminum block that ensured the FCFDv0 chip was kept at a constant temperature throughout the testing phase. 
The output data from the FCFDv0 chip was recorded using a Lecroy Waverunner 8208HD oscilloscope with a bandwidth of 2~GHz and sampling rate of 10~Gsps. 

\subsection{Laser and Beta Source Setup}

\begin{figure}[htp]
  \begin{center}
    \includegraphics[width=0.85\textwidth]{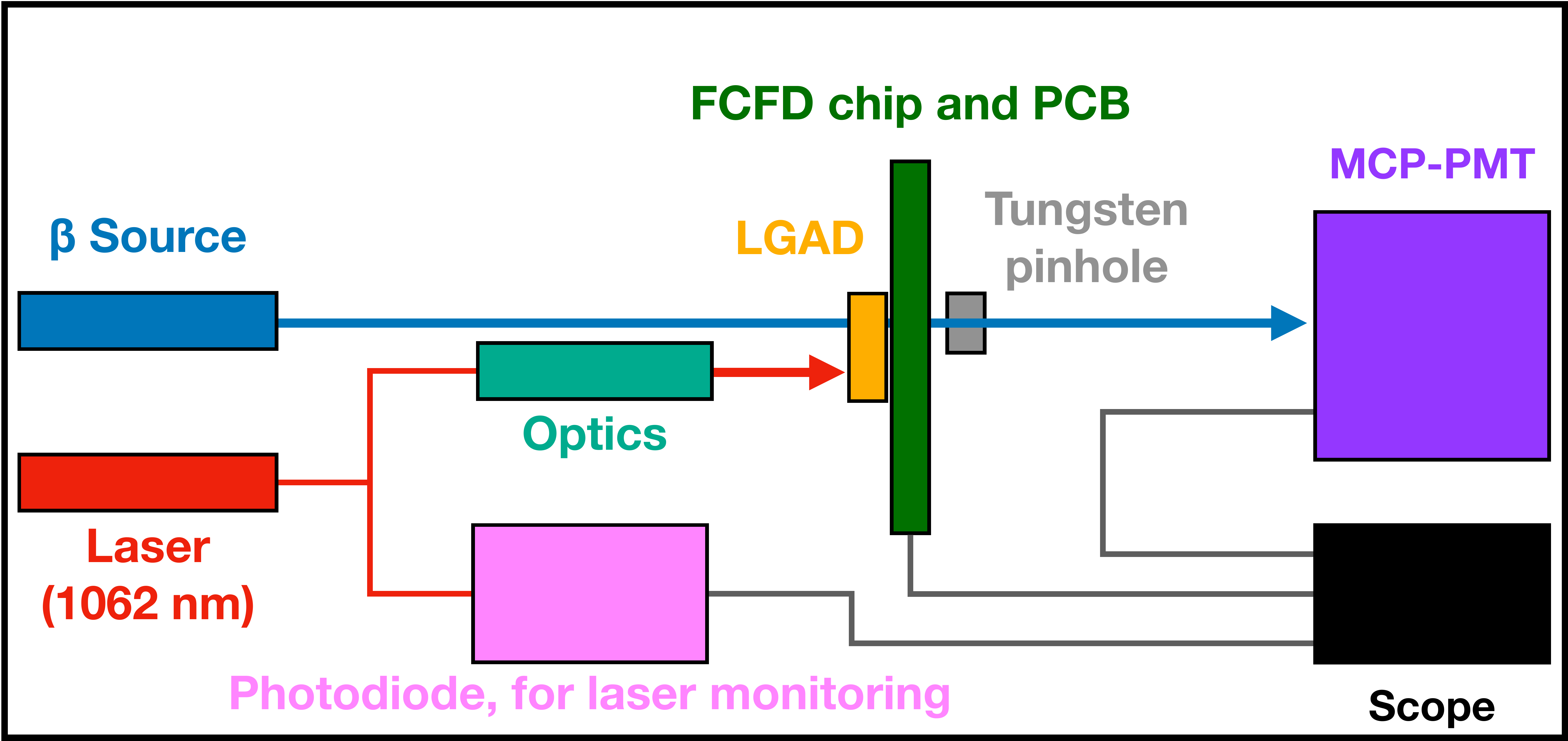}\\
  \caption{Schematic diagram of the LGAD sensor and FCFDv0 readout board within the environmental chamber housing both the laser and beta particle source. }
  \label{fig:LaserBetaSetupSchematic}
  \end{center}
\end{figure}

\begin{figure}[htp]
  \begin{center}
  \includegraphics[width=0.85\textwidth]{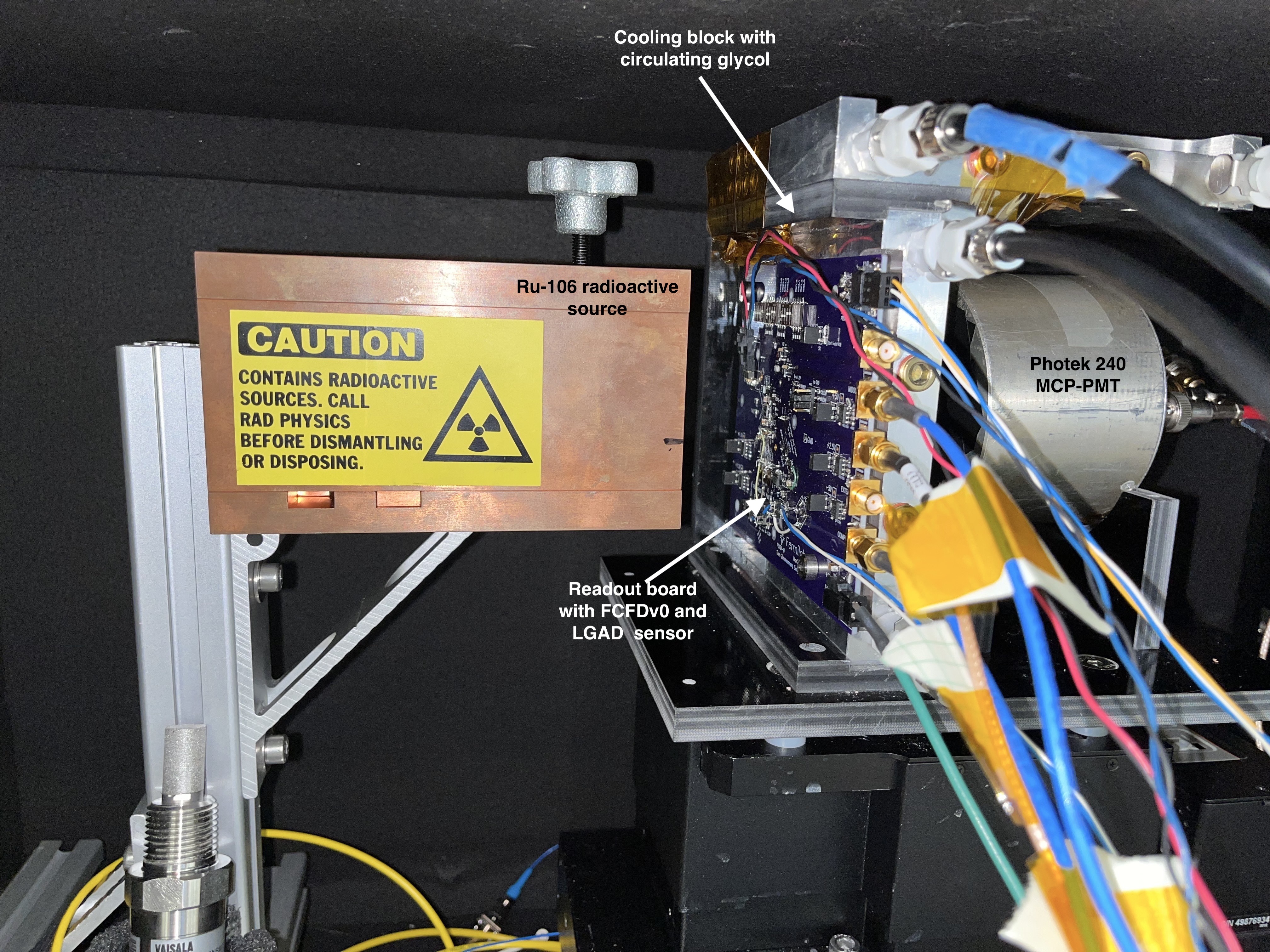}\\
  \caption{Experimental setup inside the environmental "dark" box. The radioactive source box, cooling block, Photek 240 MCP-PMT and the FCFDv0 readout board are indicated in the photograph.}
  \label{fig:LaserBetaSetupPhoto}
  \end{center}
\end{figure}

A schematic diagram of the important components of the experimental setup for the laser and beta source measurements is shown in Figure~\ref{fig:LaserBetaSetupSchematic}.
The readout board containing the LGAD sensor and the FCFDv0 readout chip is mounted onto a custom aluminum cooling block and a movable motorized stage. 
A water-glycol solution was circulated through the cooling block to maintain a constant temperature of 20 degrees Celsius.
As shown in Figure~\ref{fig:LaserBetaSetupSchematic}, we mounted two alternative sources of potential signals.
The first type of signal is a picosecond pulsed laser from Advanced Laser Diode Systems at a wavelength of \SI{1062}{\nm} mounted onto a fixed structure. 
The laser has a pulse width of 32~ps and was operated at a pulse rate of 1~kHz.
The intensity is tunable up to a maximum power of 50~$\mu$W.
The laser is split and $90\%$ of the light is directed onto a photodiode used to monitor the laser light output for stability, while the remaining $10\%$ is delivered to the LGAD. 
The second type of signal is a Ruthenium 106 beta source with an activity of about \SI{5}{mCi}.
The source is enclosed in a shielded box with a circular opening at the front with a diameter of \SI{3}{\cm}. 
To control the trajectory of the emitted beta particles, a tungsten collimator with an opening of diameter \SI{0.5}{\mm} is placed behind the LGAD sensor and a Photek 240 micro-channel plate photomultiplier (MCP-PMT) is placed behind the collimator at a distance of about \SI{5}{\cm} to detect the beta particle that pass through the LGAD.  
The MCP-PMT is used to measure the reference time-stamp of the beta particle and has a time resolution of about \SI{15}{\ps} for beta particles.
The motorized stage is used to align the beta source with the collimator and a maximum rate of coincidence between the LGAD sensor and the MCP-PMT of about \SI{2}{\Hz} was achieved. 
A photograph showing the placement of the beta source box and the readout board is shown in Figure~\ref{fig:LaserBetaSetupPhoto}.

\subsection{Beam test setup}
\begin{figure}[htp]
\centering
\includegraphics[width=0.99\textwidth]{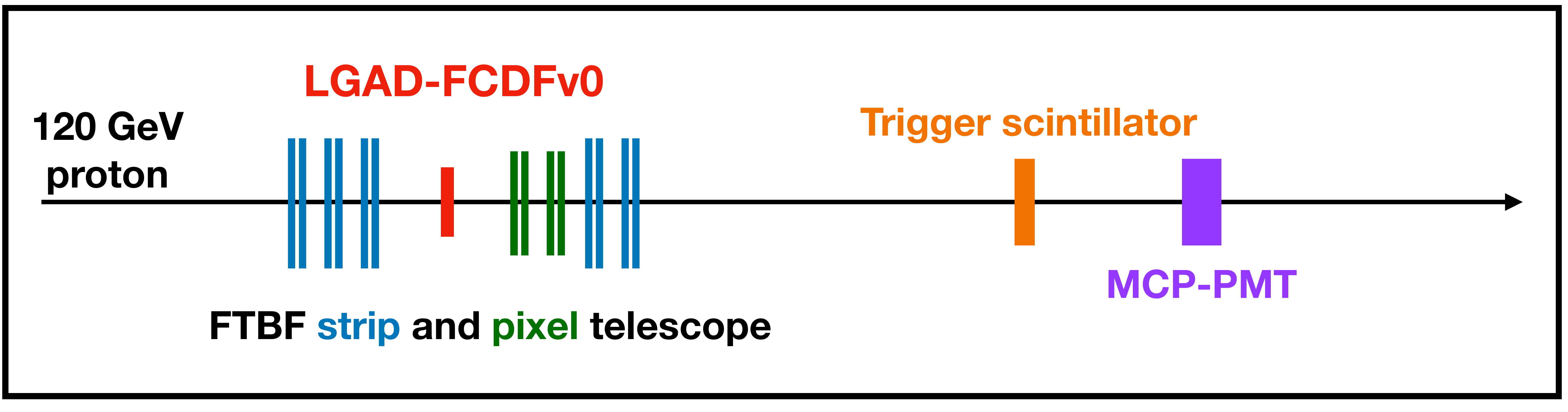}

\vspace{0.5cm}

\includegraphics[width=0.99\textwidth]{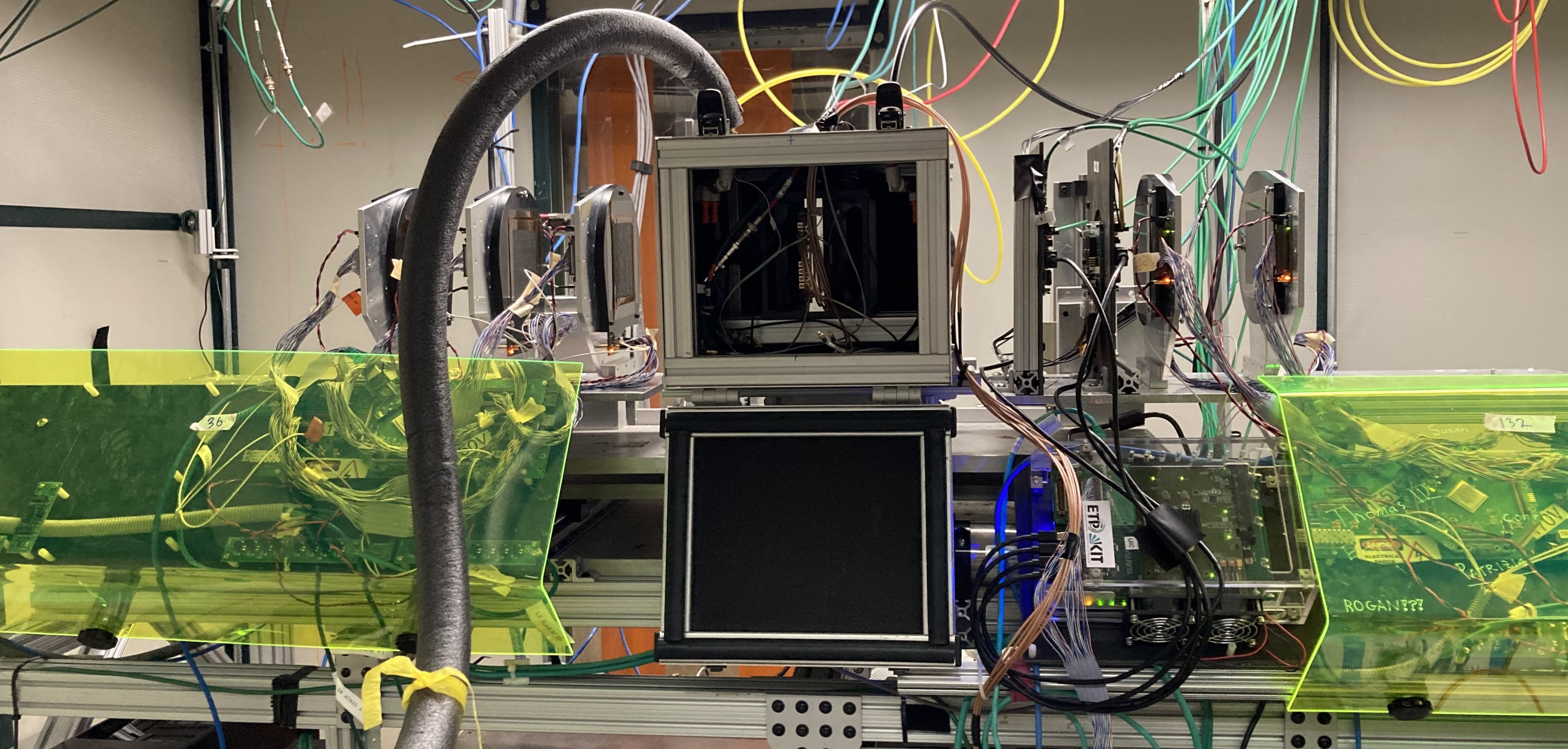}
\caption{Diagram of the FCFDv0 readout board and reference instruments along the beamline (top). The environmental chamber placed within the FTBF silicon telescope (bottom). The telescope comprises five pairs of orthogonal strip layers and two pairs of pixel layers, for a total of up to 14 hits per track. 
\label{fig:FTBF_Box}}
\end{figure}

The beam test results were collected at the Fermilab \SI{120}{\GeV} proton test beam facility (FTBF), using similar LGAD characterization setups described in previous results~\cite{Apresyan:2020ipp,Heller_2022,Madrid:2022rqw}. 
The experimental setup includes a silicon tracking telescope for impact position measurements of each proton, and the same Photek 240 MCP-PMT described above to obtain the reference timestamp for time resolution measurements. The MCP-PMT response has a resolution of about \SI{10}{\pico \s} for \SI{120}{\GeV} protons.
The FCFDv0 and MCP-PMT waveforms were recorded using the eight channel Lecroy Waverunner 8208HD oscilloscope mentioned above. 
The FCFDv0 readout board is mounted on an aluminum cooling block with temperature maintained at 20 degrees Celsius. 
Figure~\ref{fig:FTBF_Box} shows a diagram of the setup along with an image of the environmental chamber that houses the FCFDv0 readout board inside the telescope.
The trigger signal is generated by a scintillator detector located downstream from the FCFDv0 board and distributed to the tracker and the oscilloscope. 
The telescope comprises five pairs of orthogonal strip layers and two pairs of pixel layers, for a total of up to 14 hits per track. 
Based on improved alignment procedures detailed in reference~\cite{Madrid:2022rqw}, position resolutions of the impact position as low as \SI{5}{\micro\m} are achieved with this setup. 

\section{Results}\label{sec:results}
We characterize the performance of the FCDFv0 chip by measuring time resolution and mean time response under various experimental scenarios including charge injection, laser light injection, beta particle source, and proton beam. 
These results collectively demonstrate a consistent and high level of performance, and represents a successful demonstration of the utility of the CFD ASIC readout chip concept.

\subsection{Charge injection}\label{subsec:Qinj}
\begin{figure}[htp]
  \begin{center}
  \includegraphics[width=0.49\textwidth]
  {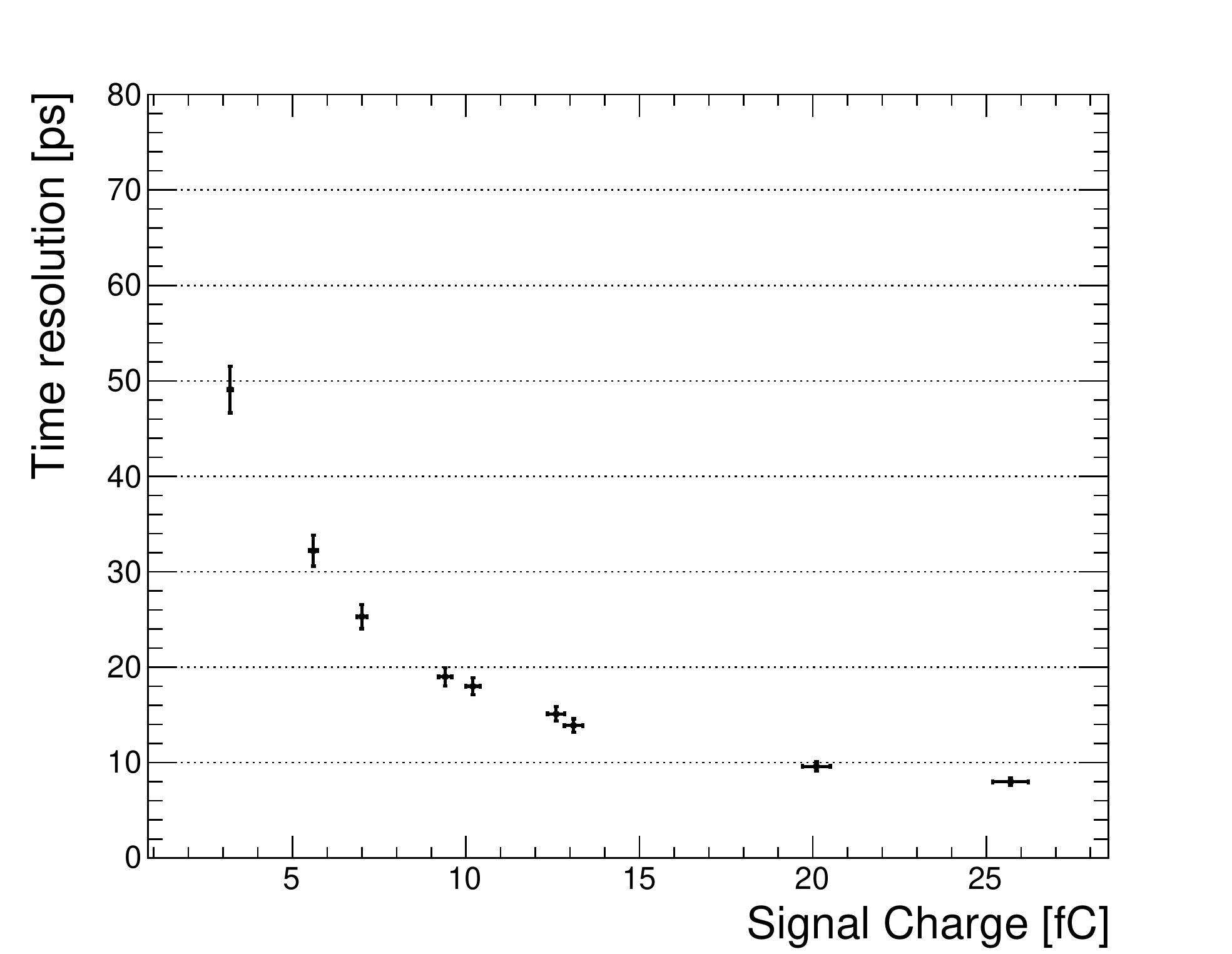}
  \includegraphics[width=0.49\textwidth]{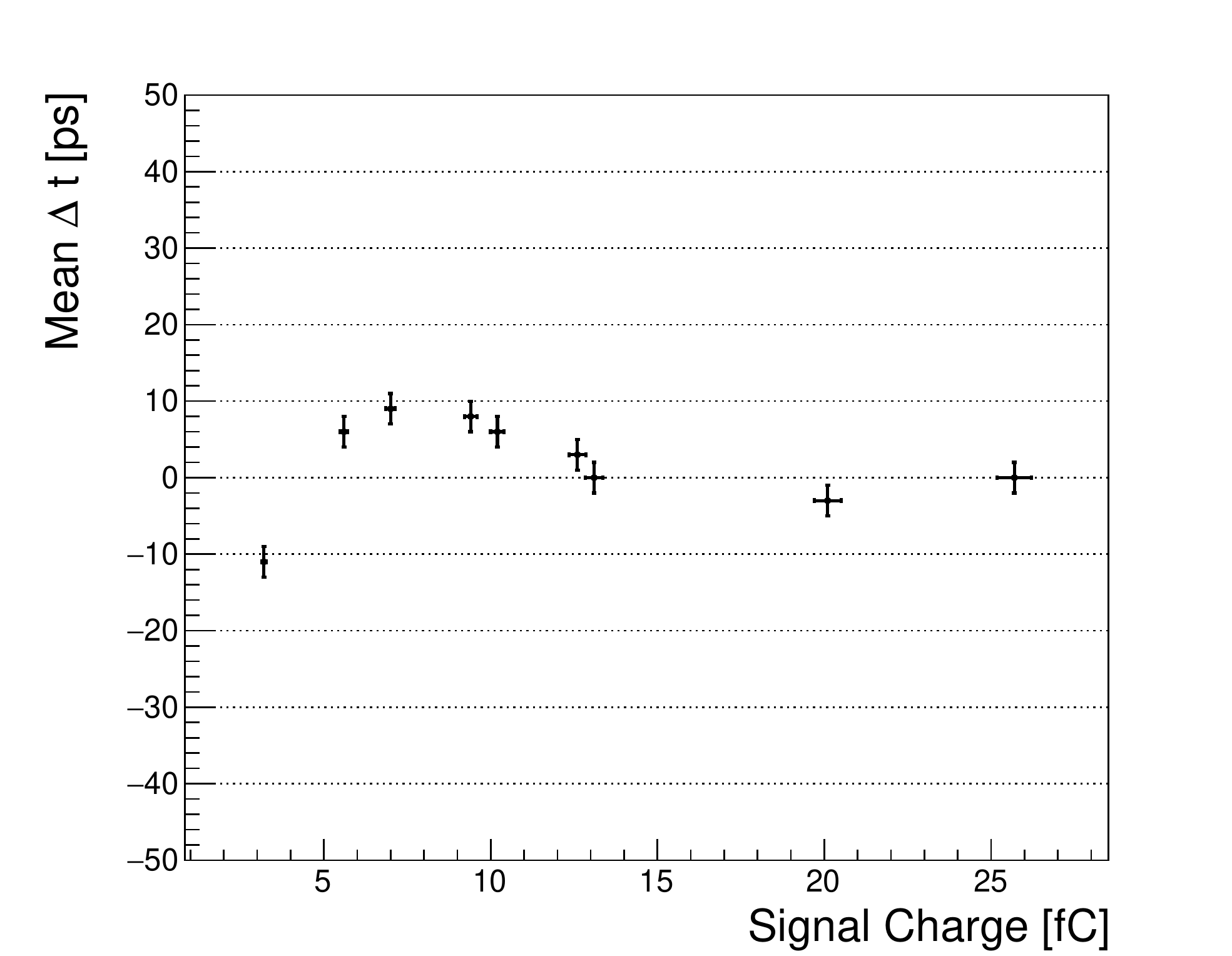}
  \caption{The time resolution (left) and mean time response (right) as a function of the
  FCFDv0 internal injected signal charge.}
  \label{fig:ChargeInjection}
  \end{center}
\end{figure}

The FCFDv0 chip was designed with a special charge injection mechanism for performance validation.
The amount of charge injected is selected through a number of switches that progressively add larger amounts of injected charge. 
The pulse shape of the injected charge is based on simulated LGAD signal pulses and is representative of minimum ionizing particle signals. 
Using this charge injection mechanism, we scan the size of the injected signal charge between 4 and \SI{26}{\femto\coulomb}, and measure the corresponding time resolution and mean time response as shown in Figure~\ref{fig:ChargeInjection}.
We obtain time resolution better than \SI{8}{\ps} at the highest injected charge.
Between 6 and \SI{26}{\femto\coulomb}, we observe a stable time response and no evidence of residual time-walk.
All mean arrival times fall within a \SI{10}{\ps} window, thus validating the effectiveness of the CFD concept. 

\subsection{Laser illumination}\label{subsec:laser}
\begin{figure}[htp]
  \begin{center}
  \includegraphics[width=0.69\textwidth]{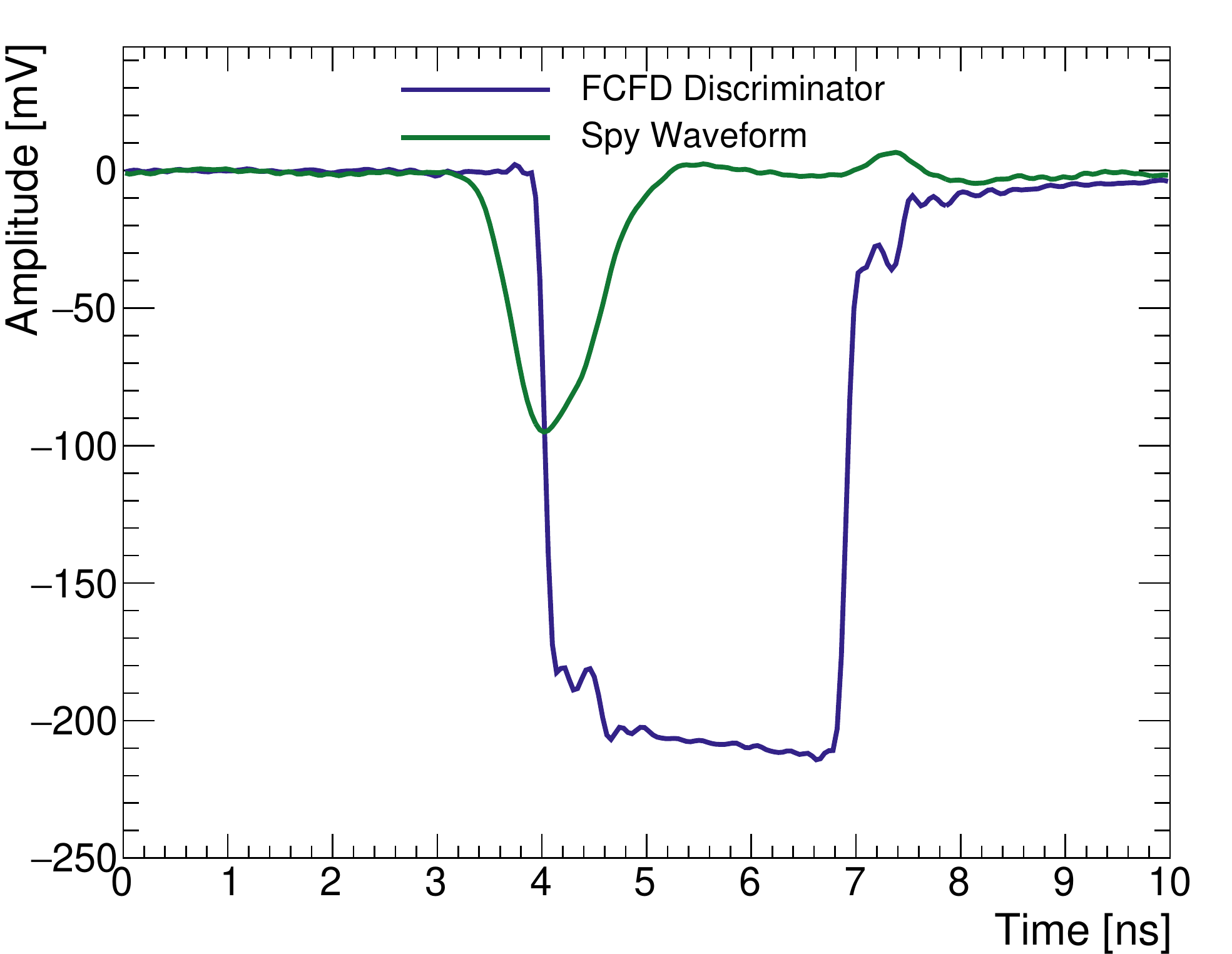}\\
  \caption{Candidate waveforms of the FCFDv0 discriminator output (blue) and the FCFDv0 input signal spy (green).}
  \label{fig:waveform}
  \end{center}
\end{figure}

\begin{figure}[htp]
  \begin{center}
  \includegraphics[width=0.49\textwidth]{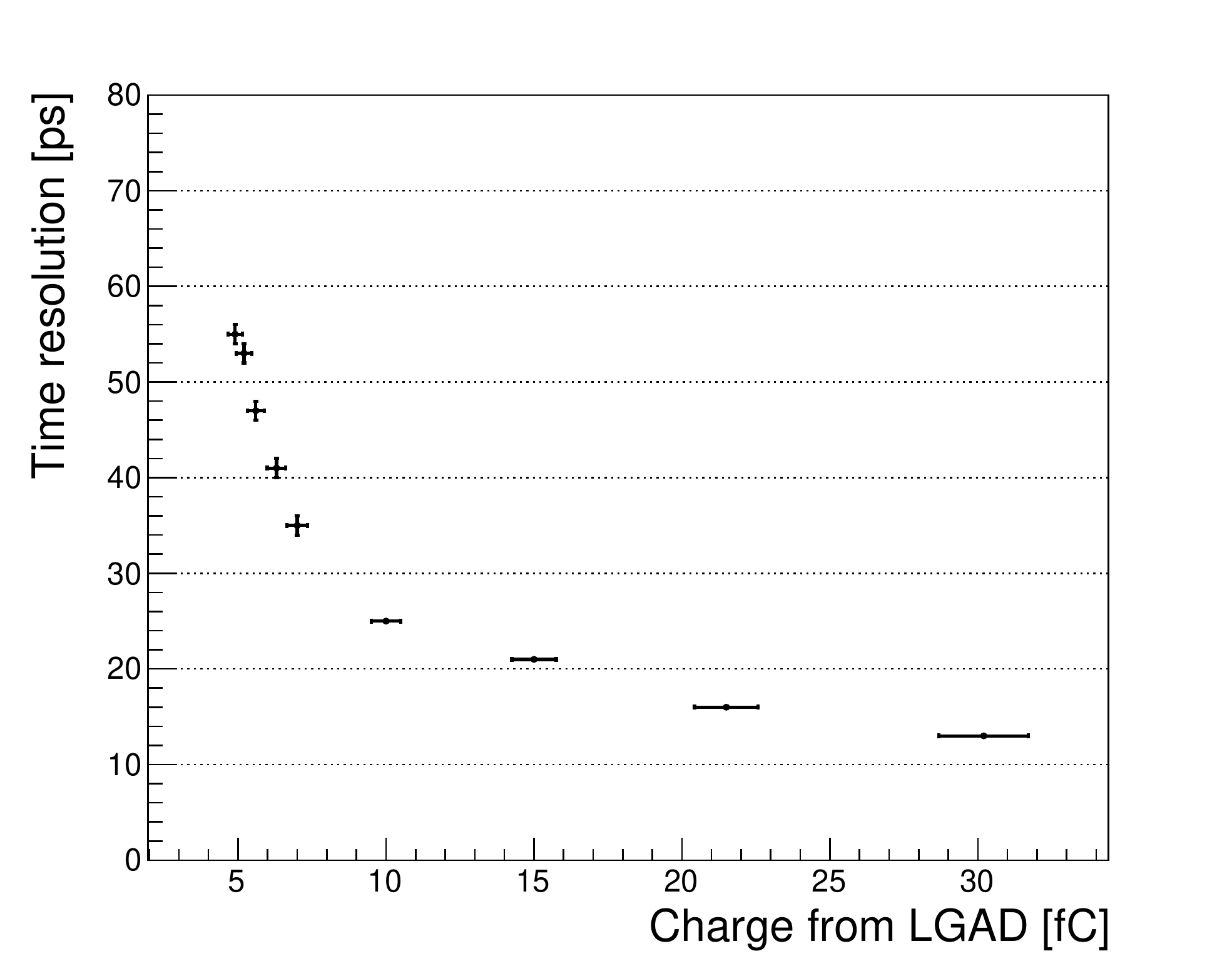}
  \includegraphics[width=0.49\textwidth]{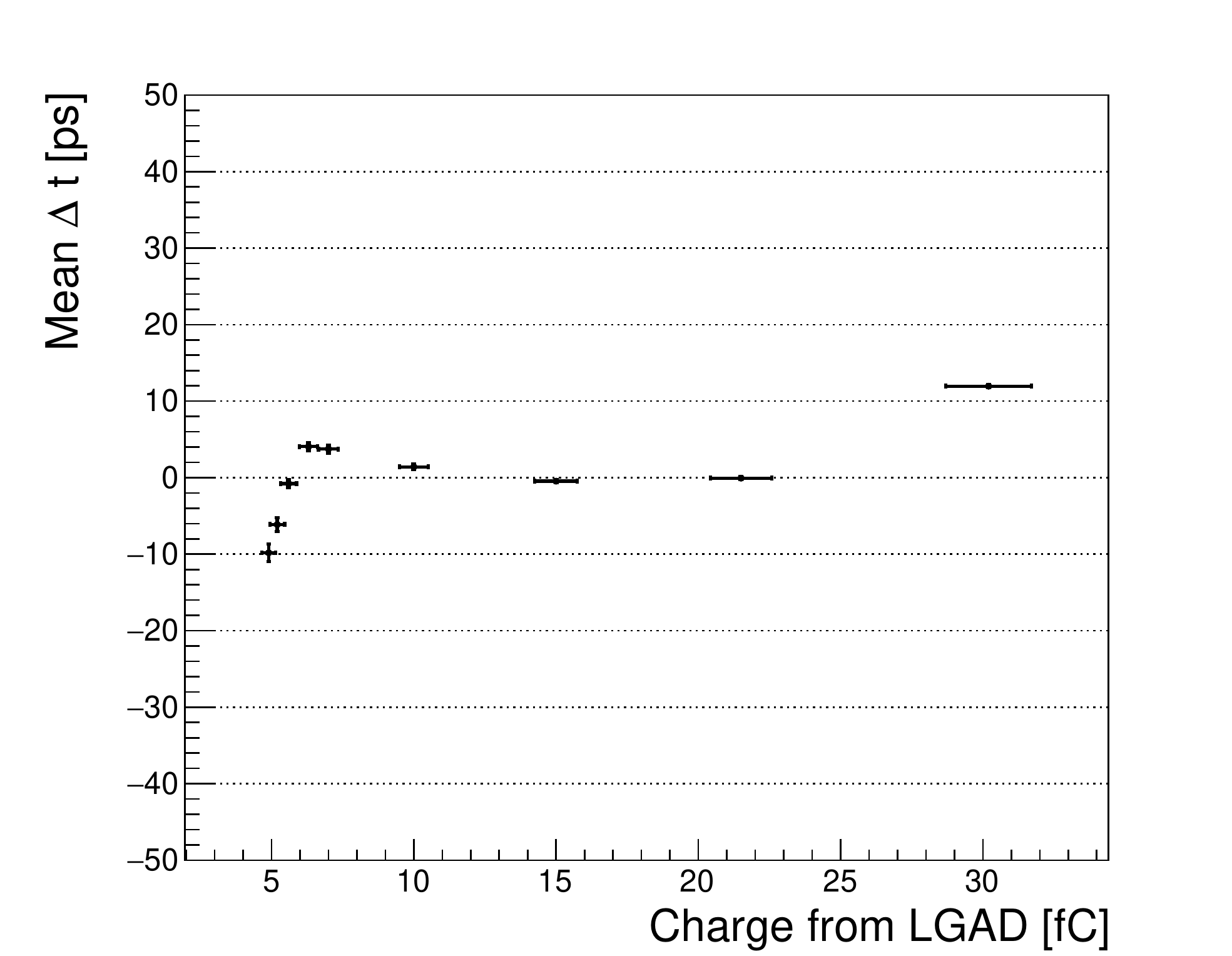}\\
  \caption{The time resolution (left) and mean time response (right) as a function of the charge of the detected laser signal. }
  \label{fig:LaserInjection}
  \end{center}
\end{figure}

With a picosecond pulsed laser, we illuminate the LGAD sensor with photons to mimic the response of a minimum-ionizing particle.
The intensity of the laser is varied to scan a range of signal sizes between 3 and \SI{40}{\femto\coulomb}.
To calibrate the laser intensity, we used the spy readout of the LGAD analog waveform described in ~\ref{sec:design}. Using the spy signal, we could measure the delivered charge and tune the laser intensity to correspond to a series of specific LGAD signal charges. Then, with the spy disabled, the timing measurements were performed at this series of calibrated laser intensities, and the results are shown in Figure~\ref{fig:LaserInjection}.
A sample event that shows both the discriminator output, and the waveform from the spy channel is shown in Figure~\ref{fig:waveform}.

Finally, we measure the time resolution and mean time response of the FCFDv0 chip as a function of the signal charge, shown in Figure~\ref{fig:LaserInjection}.
We observe time resolutions similar to the charge injection measurement.
For charge above \SI{30}{\femto\coulomb}, beyond the design specifications, the FCFDv0 chip begins to saturate and the time resolution becomes increasingly worse for larger signals. 
The range of signal charge appropriate for the chip will be enlarged in the next version of the chip to accommodate larger signals.
The mean time response is also observed to fall within a \SI{10}{\ps} window for charge above \SI{6}{\femto\coulomb}, consistent with the charge injection measurement. 

\subsection{Beta source}\label{subsec:source}

\begin{figure}[htp]
  \begin{center}
  \includegraphics[width=0.69\textwidth]{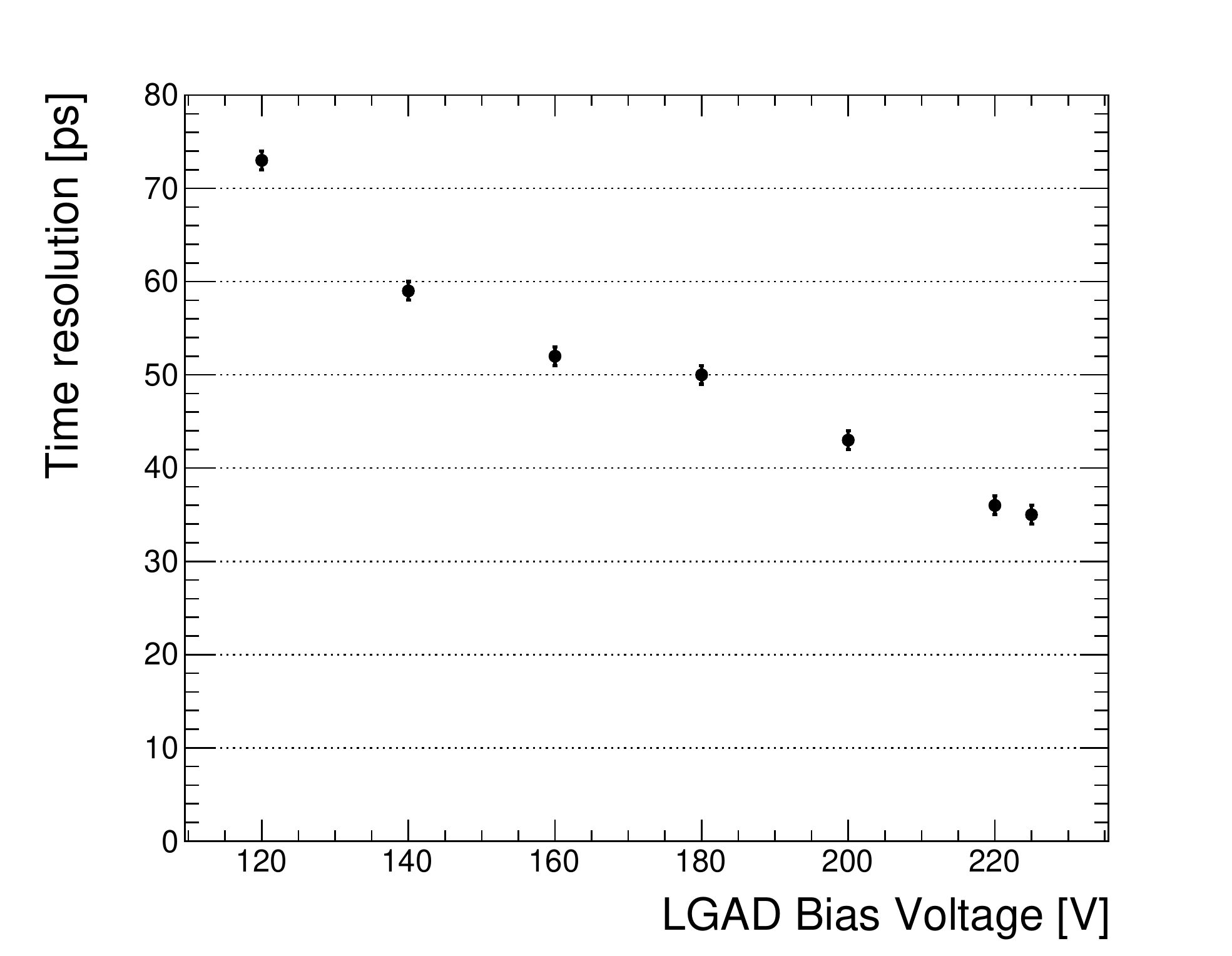}
  \caption{Time resolution measured in experiments with beta-source, shown as a function of the  bias voltage applied on the LGAD sensor. }
  \label{fig:TimeResolutionBetaSource}
  \end{center}
\end{figure}

The Ruthenium 106 source emits beta particles with energy of about 2~MeV. As demonstrated in ~\cite{Heller_2022}, the population of beta particles that pass through the tungsten pinhole and reach the MCP-PMT behind are a good approximation of minimum ionizing particles. The pinhole also truncates variation in path length between the LGAD and the MCP-PMT that would otherwise inflate the observed time resolution.
We measure the time difference between the LGAD signals from the FCFDv0 chip and the signal detected by the MCP-PMT. 
The time resolution is obtained as the sigma parameter of a gaussian fit to the time difference distribution.
In Figure~\ref{fig:TimeResolutionBetaSource} we show the measured time resolution as a function of the bias voltage applied to the LGAD sensor.
At the nominal operating bias voltage of 220~V, we measure a time resolution of about 40~ps.
Accounting for the intrinsic time jitter due to Landau fluctuations in the LGAD sensor itself and the time resolution of the MCP-PMT reference detector, this result implies a time resolution of the FCFDv0 chip that is consistent with the charge injection and laser signal injection measurements. 

\subsection{Proton beam}\label{subsec:testbeam}
\begin{figure}[htp]
  \begin{center}
  \includegraphics[width=0.49\textwidth]{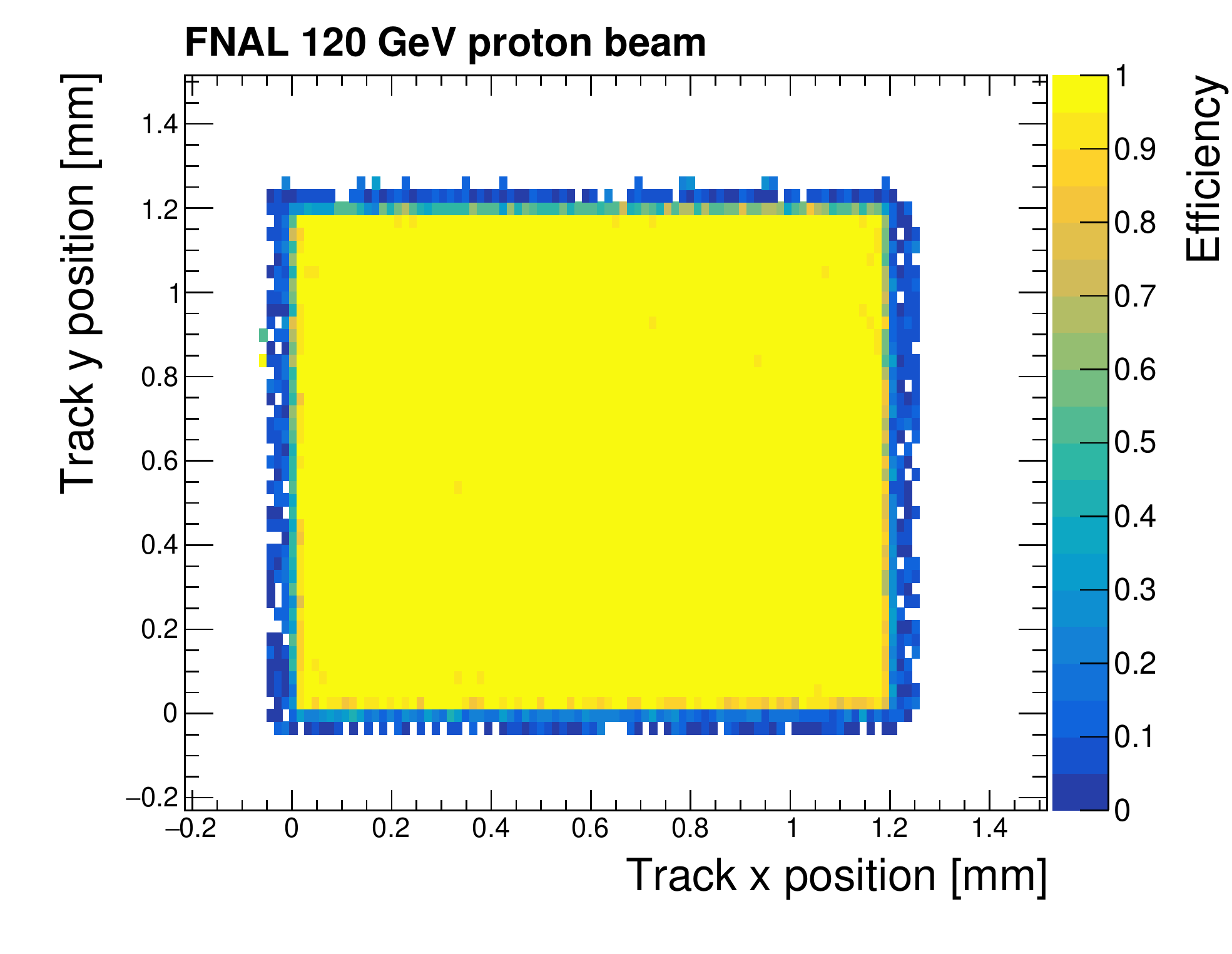}
  \includegraphics[width=0.49\textwidth]{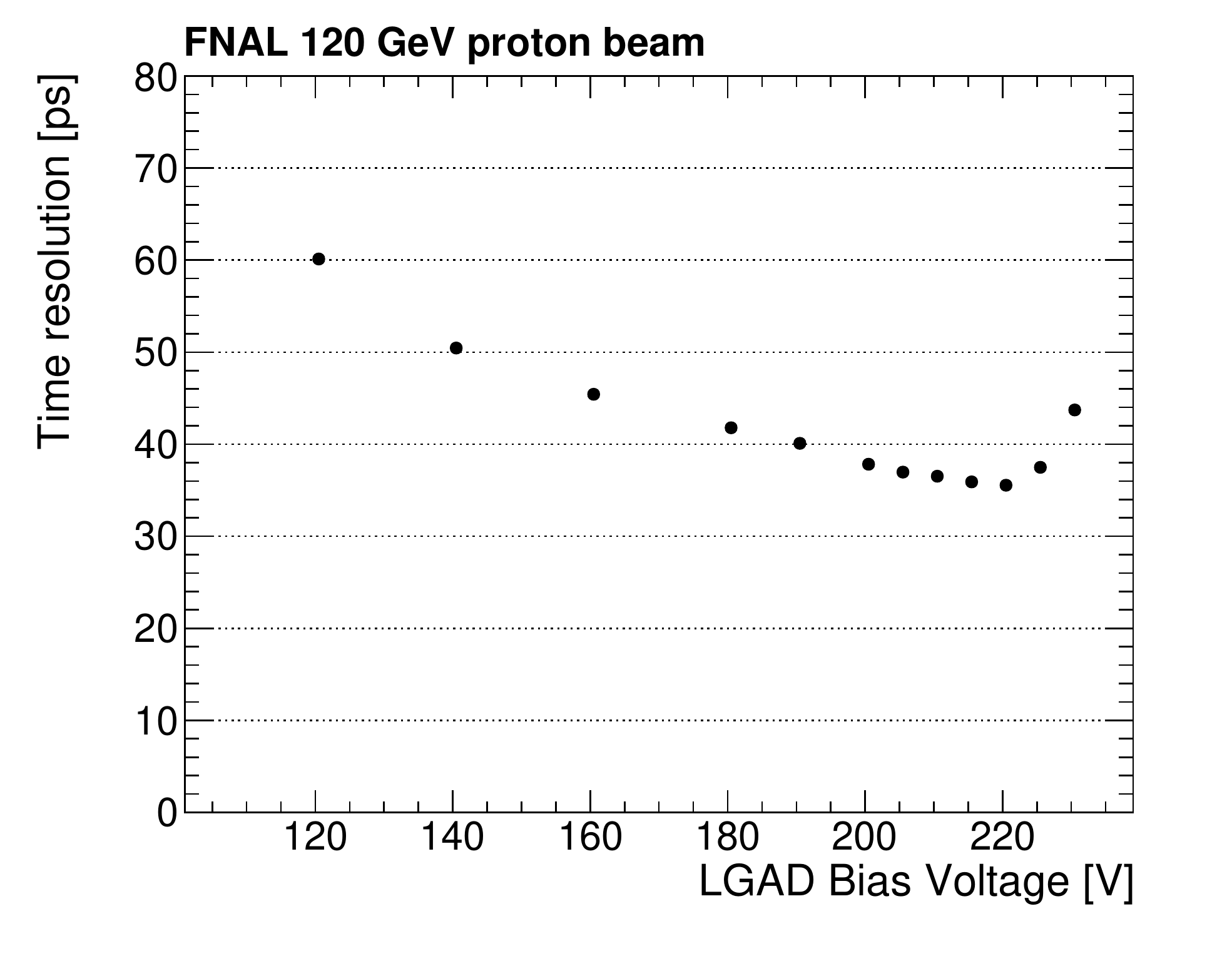}
  \caption{Efficiency as a function of the proton impact position along the horizontal (x) and vertical (y) positions measured using the test beam telescope (left). Time resolution measured as a function of the bias voltage applied to the LGAD sensor for \SI{120}{\GeV} protons (right).}
  \label{fig:Testbeam}
  \end{center}
\end{figure}

Using the FTBF facilities, we measure the performance of the FCFDv0 chip for 120~GeV protons.
On the left of Figure~\ref{fig:Testbeam}, we show measurement of the efficiency for the FCFDv0 chip to produce a valid signal as a function of the proton impact position on the LGAD sensor. 
We observe an excellent signal efficiency across the surface of the sensors showing the FCFDv0 chip was able readout all signals sent by the LGAD at a trigger rate of about 25~kHz.
On the right of Figure~\ref{fig:Testbeam}, we measure the time resolution as a function of the bias voltage applied on the LGAD sensor.
At the nominal operating bias voltage of 220~V, we measure a time resolution of about 36~ps, similar to the beta source measurement.
Accounting for the impact of the intrinsic time jitter due to Landau fluctuations in the LGAD sensor and the time resolution of the MCP-PMT reference detector, this result again implies a time resolution of the FCFDv0 chip consistent with all previous measurements presented, including the the charge injection, laser signal injection, and beta source measurements. 

\section{Conclusions and Outlook}\label{sec:discussion}

The development of a readout ASIC for LGAD sensors using the constant fraction discriminator technique is an important technological milestone for the development of future 4D tracking detectors with precision timing capabilities.
The calibration and monitoring necessary for achieving the ultimate timing precision for 4D tracking detectors with millions of channels presents an overwhelming challenge. 
In this paper, we presented measurements of the performance of the first such readout ASIC, the FCFDv0 chip, using four different experimental setups including charge injection, laser illumination, beta particle source, and proton beam. 
Measurements of the performance of the FCFDv0 chip in all four experimental setups are found to be consistent and indicate time resolution better than 10 ps for signals with charge above 20~fC. 
Measurements of the mean time response also indicate that the measured time is independent of the signal size and suggests that explicit time-walk corrections are no longer necessary.

Short term improvements for the development of the FCFD chip includes the optimization of the saturation thresholds for the chip, the expansion of its dynamic range to cover the full range of possible signals from MIPs, and to implement a concurrent signal amplitude readout and digitization necessary for strip-based AC-coupled LGAD sensors.
Future integration of the FCFD readout chip with a digital processing and communication component would allow the construction of small-scale 4D tracking detector prototypes that could be implemented as a 4D tracking telescope for beamlines such as the one available at FTBF. 

\section*{Acknowledgements}

We thank the Fermilab accelerator and FTBF personnel for the excellent performance of the accelerator and support of the test beam facility, in particular M.~Kiburg, E.~Niner, N.~Pastika, E.~Schmidt and T.~Nebel. 
We also thank the SiDet department, in particular M.~Jonas and H.~Gonzalez, for preparing the readout board by mounting and wirebonding the LGAD sensor along with the FCFDv0 ASIC. 
Finally, we thank L.~Uplegger for developing and maintaining the telescope tracking system.

This document was prepared using the resources of the Fermi National Accelerator Laboratory (Fermilab), a U.S. Department of Energy, Office of Science, HEP User Facility. 
Fermilab is managed by Fermi Research Alliance, LLC (FRA), acting under Contract No. DE-AC02-07CH11359.
This work was also supported by the U.S. Department of Energy (DOE), Office of Science, Office of High Energy Physics Early Career Research program award. 



\bibliographystyle{elsarticle-num} 
\bibliography{main}{}


\end{document}